# Cometary panspermia explains the red rain of Kerala

Godfrey Louis & A. Santhosh Kumar

School of Pure and Applied Physics, Mahatma Gandhi University,
Kottayam – 686560, Kerala, India.
E-mail: godfreylouis@vsnl.com

Date: October 5, 2003

**Red coloured rain occurred in many places of Kerala in India during July to September 2001 due to the mixing of huge quantity of microscopic red cells in the rainwater. Considering its correlation with a meteor airbust event, this phenomenon raised an extraordinary question whether the cells are extraterrestrial. Here we show how the observed features of the red rain phenomenon can be explained by considering the fragmentation and atmospheric disintegration of a fragile cometary body that presumably contains a dense collection of red cells. Slow settling of cells in the stratosphere explains the continuation of the phenomenon for two months. The red cells under study appear to be the resting spores of an extremophilic microorganism. Possible presence of these cells in the interstellar clouds is speculated from its similarity in UV absorption with the 217.5 nm UV extinction feature of interstellar clouds.**

## 1. Introduction

Panspermia, the theory that the seeds of life are every where in the Universe has been gaining more support recently on the basis of several new findings. Modern version of panspermia considers comets as the delivery vehicles that spread life throughout a galaxy [1-3]. Comets can protect cells from UV and cosmic radiation damage and comets can drop cells high in the atmosphere to float gently down [4]. Paleogeochemical evidence show that life appeared on Earth as early as 3,800 million years ago [5] or even before that [6], immediately following the Earths surface cooling. This gives too short a time for the evolution of life to take place from simple precursor molecules to the level of prokaryotic and photoautotropic cells and it leads to the argument that life has earlier originated elsewhere and then it was transported to primitive Earth [7]. There is evidence to show that microbial life can remain in a resting phase for millions of years, which can enable them to make long space travel [8-11]. There is also the possibility of liquid water in comets, which could support active life in comets [12-15]. Some of the observational data from comets have also been interpreted as evidence to prove biological content in comets [16]. In the extreme conditions in comets, if not in active state, life can be expected to be present as spores. Spores in the dormant state, undergo no detectable metabolism and exhibit a high degree of resistance to inactivation by various physical insults [17]. Thus the most possible means by which microorganisms can arrive in a planet after a journey in space must be as spores. Considering the universal nature of biochemistry [18], the chemical makeup of extraterrestrial life forms can be expected to be similar to the one found on Earth.

Recently there have been a few claims of finding extraterrestrial life. McKey *et al.* [19] have found structures similar to microfossils of nanobacteria in a Martian meteorite, which was interpreted as evidence for life in Mars. To test the idea of cometary panspermia, Narliker *et al.* [20] have performed a stratospheric sample collection experiment using a balloon and found microorganisms in the air samples collected over Hyderabad in India at various heights up to 41 km. Wickramasinghe *et al.* [21, 22], argue that these microorganisms are of extraterrestrial origin and consider this finding as evidence to vindicate the idea of cometary panspermia.

In this paper we open a new finding in support of cometary panspermia. A study of the red rain phenomenon show that the microscopic cells that coloured the rainwater originated most possibly from a cometary meteor that disintegrated in the upper atmosphere above Kerala on 25th July 2001. A physical study of the cells indicate that the cells are spores of an extremophilic microorganism and hence we argue that the red rain phenomenon of Kerala is a case of cometary panspermia and the red cells are the first clear example of life beyond Earth. Though this claim is extraordinary, there appears no other less extraordinary way to explain the mystery of red rain in Kerala.

## 2. The red rain Phenomenon

The mysterious red rain phenomenon occurred over different parts of Kerala, a State in India. The news reports of this phenomenon appeared in *Nature* [23] and various newspapers and other media and are currently carried by several websites [24-29]. The red coloured rain first occurred at Changanacherry in Kottayam district on 25th July 2001 and continued to occur with diminishing frequency in Kottayam and other places in Kerala for about two months. The red colouring of the rainwater was found to be entirely due to the presence of tiny red cells about 10 micrometers in size, which appeared dispersed in rainwater. These cells had some similarity in appearance with alga cells. From the magnitude of the phenomenon, it can be estimated that several thousands of kilograms of these cells are required to be there in the atmosphere to account for all the red rain. From where the huge quantities of these cells originate and how they reach the rain clouds to cause red rain for two months is found to be a mystery. In majority of cases the colour of the rain was red. There were a few cases of yellow rain and rare unconfirmed cases of other colours like black, green, grey etc. Coloured hailstones were another reported case. It is easy and non-controversial to dismiss this phenomenon without much study by stating some conventional, simple and unproved reasons like: dust from Sahara, pollen grains, volcanic dust from distant volcanoes, fungal spores from trees, algae from sea and factory pollution etc. But a closer examination of the features of this phenomenon and the properties of the cells show that these kinds of reasons are not valid.

A study of the distribution of the red rain incidences with location and time was done using the data available on this phenomenon. This data was mostly compiled from the reports that appeared in local leading Malayalam language newspapers, which have an extensive network of reporters covering all parts of Kerala. In many cases photographs of the collected rainwater were given with the news item. Being an unusual phenomenon the local press have given much importance to this. Still there can be several cases where people have not reported the incidence to the press. Also there can be several cases, which went unobserved by the people, such as the cases, which occurred during night. But the available data is sufficient to show the trend and nature of the phenomenon (See supplementary information for a list of red rain incidences with time and place).

A plot (Fig.1a) of the number of coloured rain incidences in Kerala on different dates shows that about 75% of the total 124 listed cases occurred during the first 10 days. A plot (Fig.1b) of the average rainfall data of Kerala enclosing the coloured rain period from 25th July to 23$^{rd}$ September 2001, demonstrates that the coloured rain started suddenly during a period of rainfall in the State. Thus the cells are not something which accumulated in the atmosphere during a dry period and washed down on a first rain. It was found that several cases of red rain phenomenon have occurred on rainy days after and during normal rains. Thus it cannot be again assumed that the red cells came from some accumulation in the lower atmosphere. The vessels kept in open space also collected red rain. Thus it is not something that is washed out from rooftops or tree leaves. It appears as if the rain clouds in some region are suddenly mixed with red cells. It may be argued that the cells arrived here, from a distant source like a desert in another

Fig. 1a

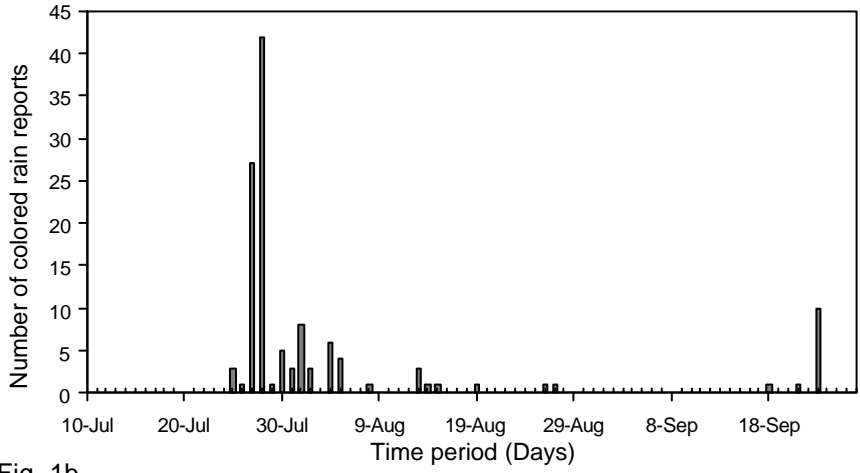

Fig. 1b

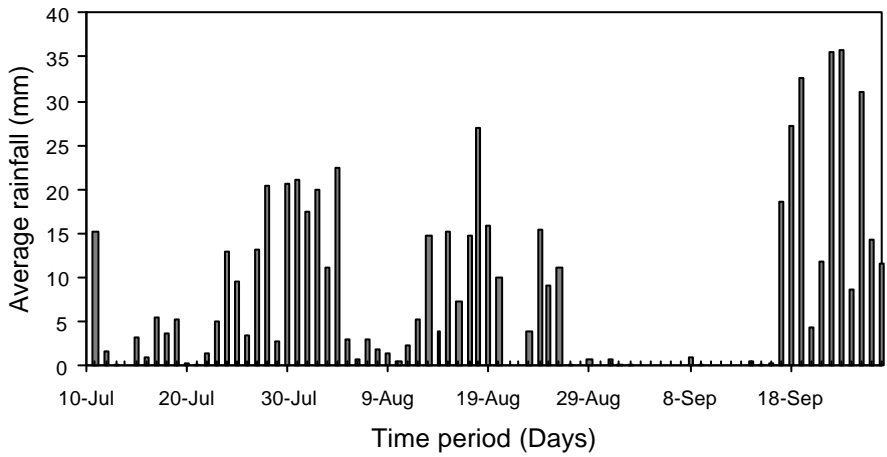

Fig. 1c

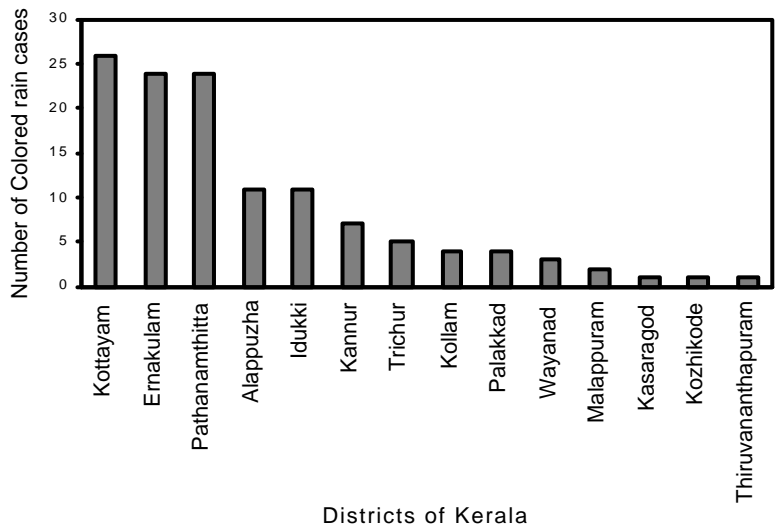

**Figure 1**. Distribution of red rain.

a) A plot of the number of coloured rain incidences in Kerala on different dates. The total number of cases reported from various parts of the state from 25$^{th}$ July 2001 to the end of September 2001 is 124. These cases are distributed in time as shown in the figure. 75% of the cases are during the first 10 days.

b) The average of the rainfall recorded in different places in Kerala from 10$^{th}$ July to 27$^{th}$ September 2001. This record shows that there was good rainfall before 25$^{th}$ July. The coloured rain phenomenon started only after the meteor airburst on 25$^{th}$ July 2001. August 27 to September 16 was a period of no rainfall and hence the absence of coloured rain reports during that period. The falling red cells during this no rain period will go unnoticed because of getting diffused in the wind of the lower atmosphere.

c) The distribution of coloured rain phenomenon in different districts of Kerala over the full period of its occurrence. The red rain first occurred at Changanacherry in Kottayam district. The maximum numbers of cases were reported from Kottayam district on the subsequent days. Ernakulam and Pathanamthitta, which also show maximum cases, are adjacent to the north and south of Kottayam respectively. Alappuzha and Idukki lie on the western and eastern sides of Kottayam respectively. This distribution is very consistent with the fall ellipse of a disintegrating meteor, which was moving from north to south and airbusted over Kottayam district on 25$^{th}$ July 2001.

part of the world, through some wind system. But in such a system it is hard to explain the repeated delivery of these cells to target over a few districts in Kerala for two months while not over other adjacent States in India, despite the changes in climatic conditions and wind pattern spanning over two months.

When the red rain reports are viewed in the background of the normal rainfall data the pattern that emerges is that of a sudden starting of red rain reports after 25$^{th}$ July 2001 and then a gradual decay of reports with time. A gap in the red rain reports is due to the absence of rainfall in the State during that period. If cell clouds are created in the stratosphere at various heights by a mechanism of meteor fragmentation and disintegration then clouds of such cells can slowly settle down to the rain clouds to give such a pattern of red rain. This idea is elaborated with some actual calculations in the next section.

The geographical distribution of the red rain cases (Fig.1c.) shows a clustering of cases in Kottayam and neighbouring districts like Pathanamthitta, Ernakulam, Idukki and Alappuzha with abrupt decrease towards the south and gradual decrease towards the north. This distribution over the geographical area can be explained by considering the path and the location of final airbust of the meteor. This idea is also elaborated in the following section.

## 3. The meteor connection

Only direct evidence available to support the idea of meteor airbust is the extreme loud explosive noise heard by people in Changanacherry in Kottayam district on 25$^{th}$ July early morning where the red rain was first observed. Even though this matter appeared in news media we have personally interviewed the people who have experienced this loud noise to understand the nature of sound they heard. The people described the sound they heard at about 5:30 a.m. on 25$^{th}$ July 2001, as an extremely loud and sharp noise which was distinctly different from thunder. Their description of the nature of the sound agrees with that of a sonic boom from a meteor [30-32]. Sonic boom is a high amplitude low frequency sound, which lasts for only a fraction of a second.  This kind of infrasonic sound can induce resonant vibrations in building structures giving the fearsome impression of a roof collapse which some of them agreed to have experienced. The sound cannot be that of a thunder because thunderstorms are characterised by a series of small and large lightning strokes and in the present case people heard only one loud bang. Thus it can be reasonably argued that the extreme loud noise heard by the people was a sonic boom and a meteor airburst did take place on 25$^{th}$ morning.  The red rain was first observed at Changanacherry a few hours later at around 8:45 a.m. on the same day. Then on the following days there were reports of red rain from other places which are separated by as distant as few hundred kilometres.

The first indirect evidence for a meteor connection emerges when one looks at the distribution of red rain incidences over the geographical area. If the final stages of meteor fall was over Kottayam and near by districts then these places should receive maximum amount of red rain. This is found to be true from the red rain's geographic distribution pattern as shown in (Fig.1c). The path of the meteor travel can also be guessed from the geographical distribution pattern. Usually during a large meteor fall, fragmentation of the original body takes place and these fragments appear to be distributed over an elliptical area [33, 34]. This region (Fig. 2) in the present case has a size of 450 km by 150 km. This large area is possibly due to the fragile nature of this cometary meteor, which causes fragmentation during its flight in the upper atmosphere at a low angle. The fragments would get disintegrated to fine particles during its downward travel. Thus each of the fragments can form a column of fine particles made up of red cells.  Each of these columns, which make slow vertical decent from the stratosphere, can finally interact with the rain clouds to cause highly localised red rain phenomenon at different places. Hence each case of red rain report can be equated to a

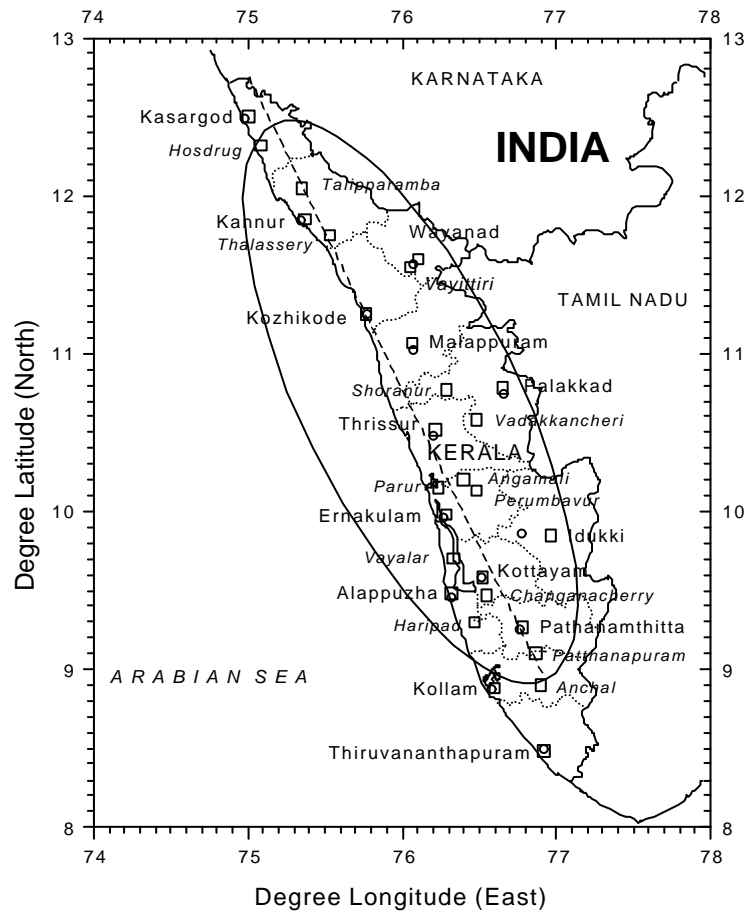

**Figure 2**. Fall ellipse of the cometary meteor.

> The direction of meteor travel and the fall ellipse, as inferred from the pattern of red rain occurrence in various places. The meteor had probably airbusted above an area between Changanacherry and Pathanamthitta while moving in the southeast direction shown by the dashed line. The fragments of the fragile meteor got disintegrated and turned into packets of particles in the atmosphere, which subsequently caused red rain and associated phenomenon by mixing with rain clouds. The fall ellipse has a size of about 450 km by 150 km and the red rain phenomenon occurred within this elliptical region. There are no reports of red rain from the neighbouring states Tamil Nadu and Karnataka. The distance per degree of latitude and longitude in Kerala region is about 110 km/deg.

fragment of the meteor. Maximum cases of red rain and associated phenomenon occurred at places in Pathanamthitta and Kottayam and Ernakulam districts (Fig. 1c), starting from the first few days of red rain phenomenon. These districts mark the head of the elliptical region. The tail region lies in the Kannur district in the north. From the above pattern of red rain, it can be inferred that while falling to the ground like a re-entering space craft the meteor has travelled from north to south in a south-east direction above Kerala as shown in Figure 2 and airbusted above Kottayam district.

Since the airbust occurred above Kottayam, only very little fragments reached the south most district Thiruvananthapuram causing little red rain there.

The second indirect evidence is the distribution of the occurrences of red rain in time. This distribution of the red rain in time can be explained using a calculation of the settling rate of the cells in the stratosphere. The dominant mechanism by which particles are removed from stratosphere is by particle settling. The settling velocity of small spherical particles is described mathematically by Stokes Law, given as:

$$v = \frac{gD^2(\rho_{part} - \rho_{fluid})}{18\mu} \qquad \text{(Eq. 1)}$$

Where v = settling velocity [m/s], g = gravitational constant [9.8 m/s$^2$], D = particle diameter [m], $\rho_{part}$ = density of particle [kg/m$^3$], $\rho_{fluid}$ = density of air [kg/m$^3$] and $\mu$ = viscosity of air [kg/m.s].

The red rain particles are of about 10 micrometers in size and assuming their density as one and half times as that of water, the settling velocity of these particles are computed using Eq. 1 and tabulated in Table 1 for various altitudes in the atmosphere. From this,

**Table 1.** Settling velocity of cells in atmosphere

| Altitude km | Air density Kg/m$^3$ | Air viscosity x10$^{-5}$ Kg/m/s | Settling velocity m/day |
|---|---|---|---|
| 00 | 1.225 | 1.789 | 394 |
| 05 | 0.73612 | 1.628 | 433 |
| 10 | 0.41271 | 1.457 | 484 |
| 20 | 0.08803 | 1.421 | 497 |
| 30 | 0.01801 | 1.476 | 478 |
| 40 | 0.00385 | 1.604 | 440 |
| 50 | 0.000097 | 1.703 | 415 |

The table lists the computed settling velocity (using Eq.1), at various altitudes in Earth's atmosphere, for spherical particles having a size of 10 micrometers and density 1500 Kg/m$^3$. This table shows that the settling velocity do not vary widely with altitude, despite the large variation in air density. The values of air density and viscosity are that of international standard atmosphere.

for the stratosphere, the average settling velocity for a red rain cell can be approximated as 0.5 km per day. This shows that, in the absence of any turbulence, such a cell can take 60 days to travel down 30 km in the stratosphere. Thus it is theoretically possible for the red rain phenomenon to occur over the same geographical region for extended periods like 60 days if the stratosphere above that region contains suspended red rain particles at different heights. Disintegrating fragile cometary meteor, which contain huge quantities loosely packed red rain cells can deposit the same in the stratosphere at different heights. Depending upon the height into which the collection of particles were thrown into the atmosphere during the disintegration of meteor fragments in the air, the time they take to reach the rain clouds can differ. This explains the temporal distribution of red rain. Above simple modelling only gives an approximate solution to the actual situation. Presence of air movement or wind, variations in particle size (size has square dependence), shape and density, clustering and adhesion of small particles to form large grains etc. can greatly increase or decrease the settling rate. Thus micro particles in the stratosphere can remain there for several months before finally reaching the ground. In the absence of rain clouds, the settling red rain causing particles will be invisible if they reach lower atmosphere, as they will get dispersed in the wind. Thus if the meteor disintegration occurred during a dry season this phenomenon would have gone unnoticed.

Above analysis explain why the red rain phenomenon can last for long period like two months or more. It explains why the red rain suddenly appeared almost simultaneously at different places in Kerala, which are separated 200 to 300 km. It explains the origin of huge quantity of cells amounting to several thousand kilograms with the assumption that a cometary body which carry such cells exists. It explains why the phenomenon did not occur in neighbouring states Karnataka and Tamil Nadu while it made repeated appearance for two months, mostly concentrated in certain districts in Kerala. But it appears impossible to find a natural terrestrial process, which can account for these features of the red rain and provide such explanations.

### 4. The red rain cells

The samples of red rain cells required for the study were obtained from widely different geographical locations separated by more than 100 kilometres from districts of Kottayam, Pathanamthitta and Ernakulam. The characteristics of the cells collected from different places were the same showing a common origin. The red rain cells appears in the rainwater as a fine suspension. This gives the water a reddish brown

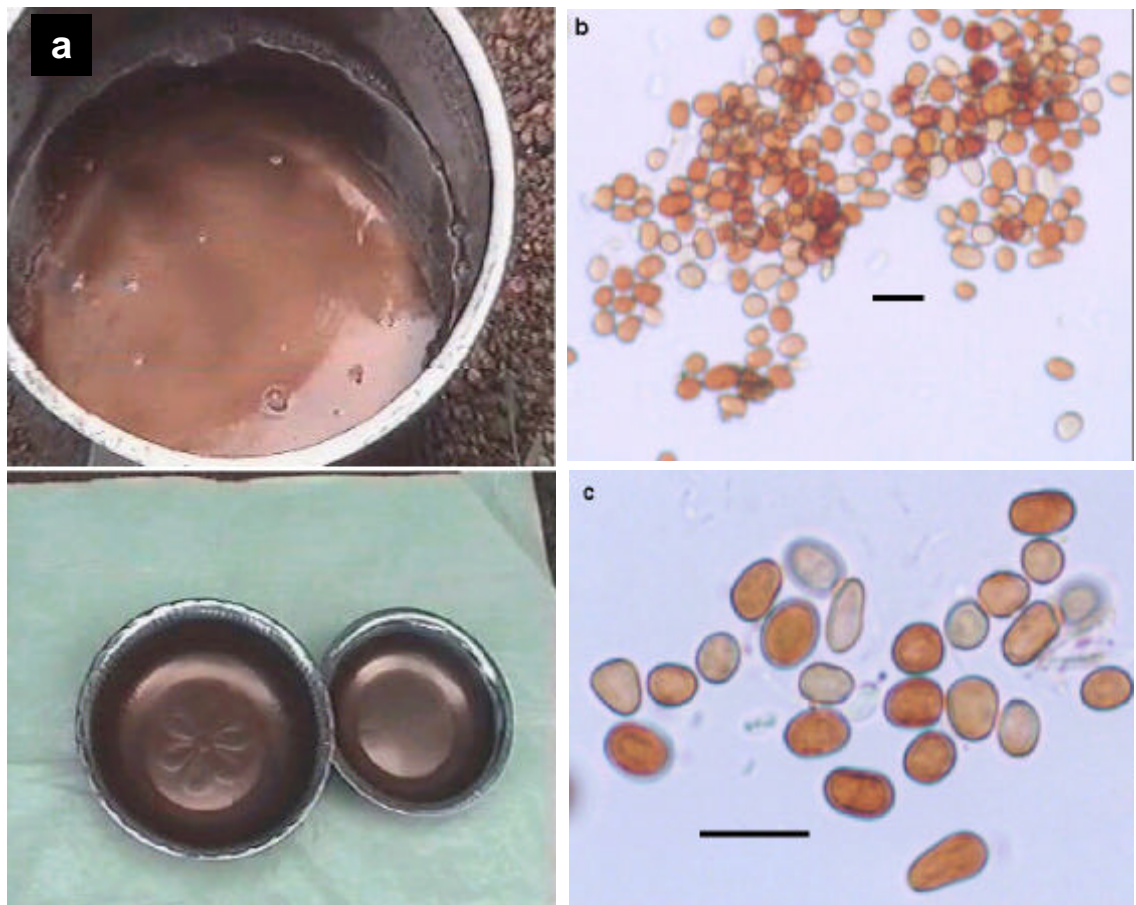

**Figure 3**. Red rainwater and cells.

a) Images of the Red Rainwater collected by some people in Pathanamthitta in their stainless steel household vessels. These images show the characteristic reddish brown colour of the red rain. An approximate calculation shows that at least 50 tones of cells are required to account for the total reported red rain cases in Kerala. This requires a cometary object of at least 10 meters in size with a very dense collection of red cells in it.

b) Photomicrograph of the red rain microorganisms under 400x magnification. Note the transparent nature of the cells. Cells at the bottom can be seen through the one on the top. Bar indicates 10 $\mu$m.

c) Photomicrograph of the red rain microorganisms under 1000x magnification. The cells show variations in shape and size. The cells show a layered structure and a cell envelope. Size varies from 4 to 12 micrometers. Bar indicates 10 $\mu$m.

colour (Fig. 3a). The cells are about 10 micrometers in size, almost transparent red in colour and are well dispersed in the rainwater. The resulting colour of the rainwater is not turbid, but it has an appearance like a red dye solution. Colour of the diluted red wine and red colour with a tint of pink under reflected light are matching descriptions for the red rainwater. When allowed to stand for several hours the cells settle to the bottom of the vessel and the water becomes colourless. Under low magnification the cells look like smooth, coloured glass beads (Fig. 3b). Under high magnifications (1000 x) their differences in size and shape can be seen (Fig. 3c). Shapes vary from spherical to ellipsoidal and slightly elongated types. The thick and coloured cell envelope of the cells can be well identified under the microscope. In a large collection only a few were found to have a broken cell envelope. No nucleus could be observed in the cells even after staining with acidified methyl green dye (Fig. 4a). The cell after the dye

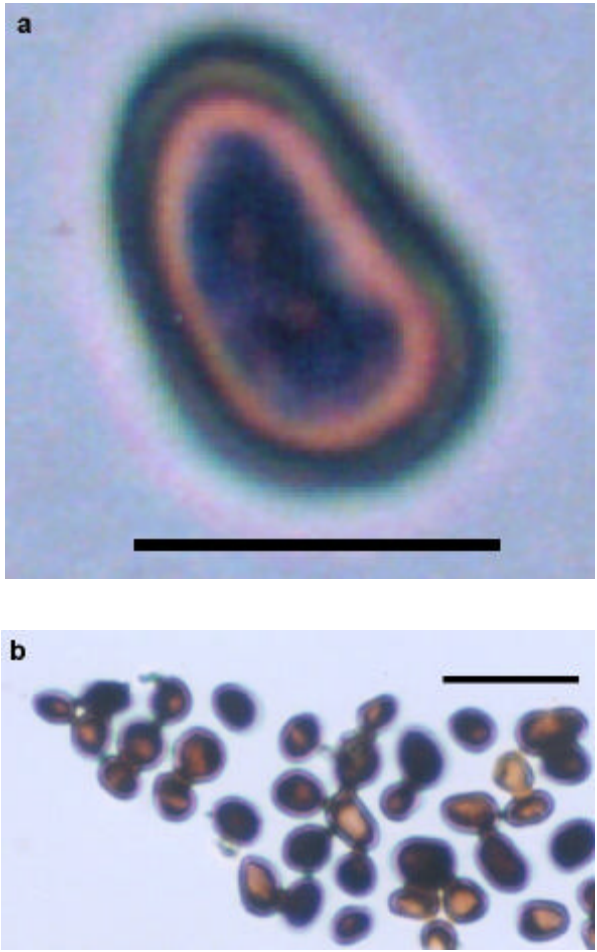

**Figure 4**. Dyed cell and heated cells.

a) Photomicrograph (1000x) of a single red rain cell which was stained by using acidified methyl green dye. Staining was achieved by keeping the cells in dye solution for few hours. For most of the cells, the dye did not penetrate beyond the first outer layer. For some cells as the one shown in figure, the dye reached the central core and produced a blue coloration there. Bar indicates 5 $\mu$m.

b) Photomicrograph (1000x) of the red cells which underwent heating to a temperature of 370° C. Water was added after heating to aid photomicrography. The size of the cells got reduced by about 20% as a result of heating. Their colour turned dark brown and the cell walls has become more visible. The black rain, which occurred at a few places, is possibly due to the presence of burnt cells, which resulted from the heating during airburst and fireball phase of the meteor. Bar indicates 10 $\mu$m.

penetration clearly shows a layered structure. The majority of the red rain cells have reddish brown colour but a small percentage of cells are white or have colours with light yellow, bluish grey and green tints. To further ascertain the biological nature of these cells, they were tested for cell proteins by Xanthoproteic test. In which the cells were heated with concentrated $HNO_3$ and then treated with NaOH solution. The yellow coloration of the solution indicated the presence of proteins in the cells.

The cells were subjected to some physical and chemical conditions to test their ability to survive extreme conditions. The cell wall is not damaged by $H_2SO_4$ but was found to get digested by concentrated $HNO_3$. The cells retained its colour and shape even after boiling in water for several minutes. To test how the cells would change on dry heating, it was placed in a heating stage under microscopic observation. As the temperature was increased progressively to a maximum of 370° C, the colour of the cells changed more and more brownish and finally blacker. No rupture or breaking of the cell wall took place, but a certain amount of shrinking in the size of the cells could be noticed (Fig. 4b). The cell wall could be seen clearly even after the heat treatment. However, the colour change and size reduction started occurring only on heating above 200° C, which suggests that these cells are possibly having an ability to survive heating while entering a planetary atmosphere. To study the effect of low pressure, the cells were placed in a vacuum chamber and the pressure was reduced to 0.01 millibar. No damage could be seen for the cells after this low-pressure treatment. This shows that the cells can survive low pressure without damage.

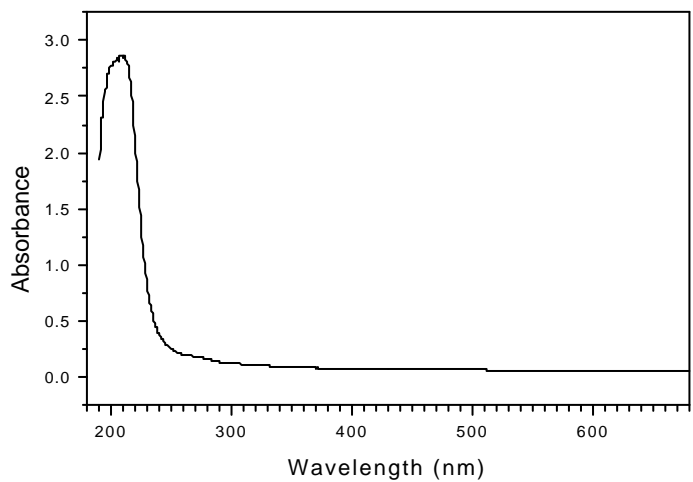

**Figure 5**. UV visible absorbance spectrum of red rainwater.

> This is the spectrum of a dilute suspension of red cells in water. A notable feature is the significant absorption in the UV region, indicating the presence of UV screening compounds in the cell, which could be a mechanism adopted by the microorganisms to escape from the damaging UV radiation in space.

The red rain was subjected to a spectrophotometric examination. The UV- Visible absorption spectrum of the cells suspended in water is shown in Fig. 5. The spectrum was recorded in comparison with distilled water using a Shimadzu model UV-2401 PC spectrophotometer. The spectrum shows significant absorption of UV radiation by the cells. This can be interpreted as an indication of the presence of effective UV screening compounds in the cells. Thus the cells can possibly survive the damaging UV radiation in space. The peak UV absorption occurs near 200 nm and this has important significance in explaining the interstellar UV extinction curve, as discussed in next section.

## 5. Discussion on cosmic origin of cells

Interstellar dust is now recognized as an important component of the interstellar medium and it is considered to play an important role in the evolution of galaxies, the formation of stars and planetary systems, and possibly, the origins of life [35]. Findings during the past few decades show that a wide range of organic compounds exist in the interstellar clouds contrary to the earlier idea that interstellar dust clouds are inorganic grains. As an extrapolation to these findings, it is now more conceivable than the past to think of the possibility of actual microorganisms in interstellar clouds and their possible arrival to Earth through comets. One of the reasons for assuming the presence of microorganisms in the interstellar clouds is the UV absorption characteristic of the interstellar clouds in which a peak is found near 217.5 nm [36]. Several explanations have been proposed by many to explain this interstellar extinction curve [37]. But none of these explanations are fully successful to explain the curve [38].

Hoyle *et al.* [39-42] have shown that the UV extinction curve is explainable by assuming the presence of microorganisms as a component of interstellar cloud and supported this claim using their experiments and theoretical considerations. Findings by some other researchers also do not rule out the possibility of microorganisms in interstellar clouds [43-45]. In this context the presence of red rain cells in interstellar space as a component of the dust clouds can be speculated on the basis of the UV absorption property of the red rain cells. The cells suspended in water shows a strong peak near 200 nm (Fig. 5). If this peak is shifted to 217.5 nm then it can account for the interstellar UV extinction. It is to be seen whether the red rain cells, when existing in interstellar space like conditions of low temperature and high vacuum, will show a shift in absorption peak to 217.5 nm. If so then the presence of red rain cells in interstellar clouds can be suspected. Comets, which originated in the interstellar clouds as aggregates of red cells and dust grains, embedded in an icy matrix of gaseous compounds can be captured by the Sun during its passage through an interstellar cloud [46]. Such comets can later seed the planets or the cells and dust that survived the formation of the solar system can also end up in comets to seed the planets. Thus cells in the interstellar clouds can reach planets as expected in cometary panspermia.

Panspermia also requires the opposite process of transport of cells from planets to interstellar clouds. In radiopanspermia single exposed microorganisms are accelerated to high velocity by stellar radiation pressure and leave a solar system and finally reach interstellar space [47]. In the present context we find that red rain cells have high UV shielding property to escape the damaging UV radiation from a star. They have larger size (3 to 4 times) when compared to bacterial spores, which means more radiation pressure on them to move away from the star. If the mass of the cells remain low due low density under dehydration, then radiation pressure can dominate over gravitational attraction. When the Sun like stars reach red-giant stage, the temperature in Earth like planets increase to high levels where only extremophilic organisms can survive. Finally when the oceans in the planets evaporate away it can possibly launch the spores of extremophilic organisms like the red rain cells into space along with the water vapour. During a red-giant phase a star emit less UV radiation, thus cause low harm to microbes in space. The cells driven out from the planet eventually reach the interstellar space and act as potential seeds to seed later planetary systems in a next generation star. Thus life can possibly continue to exist through several generations of stars.

In this paper we have shown how the red biological cells that caused the red rain phenomenon of Kerala may have originated from a cometary body. As further supporting work we have investigated the growth conditions and the reproduction techniques of the red rain extremophiles. We will be reporting immediately the extraordinary results of this study in another paper.

**Acknowledgements**

G. L. is the main author and is responsible for the facts and claims presented in this paper. A. S. K. has contributed to the paper by giving valuable assistance to G. L. for this investigation. We greatly acknowledge the help of George Varughese for collecting many of the samples and important information about the phenomenon. We thank Dr. A. M. Thomas for first approaching us with a coloured rainwater sample and Dr. Sabu Thomas for providing photomicrography facility. We also thank the teaching, administrative and technical staff members, research scholars and students of SPAP and several others who have encouraged and helped this work.

**Supplementary Information**

A List of 124 reports of colored rain cases in Kerala during the period July –September 2001, compiled from various news reports and other sources (List sorted by district).

| No. | District | Place | Location | Date | Approximate time | Available description |
|---|---|---|---|---|---|---|
| 1 | Alappuzha | Koyana | | 7/27/01 | bet. 8&10 a m | Red |
| 2 | Alappuzha | Chengannur | | 7/27/01 | bet. 8&10 a m | Red |
| 3 | Alappuzha | Pandanadu | | 7/27/01 | bet. 8&10 a m | Red |
| 4 | Alappuzha | Mulakkuzha | | 7/27/01 | bet. 8&10 a m | Red |
| 5 | Alappuzha | Pavukkara | | 7/27/01 | bet. 8&10 a m | Red |
| 6 | Alappuzha | Kuttamperur | | 7/27/01 | bet. 8&10 a m | Red |
| 7 | Alappuzha | Aroor | | 7/27/01 | bet. 8&10 a m | Red |
| 8 | Alappuzha | Haripad | | 7/27/01 | bet. 8&10 a m | Red |
| 9 | Alappuzha | Harippad | Thulamparambu | 7/28/01 | | Red |
| 10 | Alappuzha | Mavelikkara | Cheukol | 7/28/01 | Afternoon | Red |
| 11 | Alappuzha | Mavelikkara | Ponnezha | 7/28/01 | Afternoon | Red |
| 12 | Ernakulam | Perumbavur | Nedunghappra | 7/27/01 | 11 to 11:10 a m | First black drops then red |
| 13 | Ernakulam | Edayar | Kaintekara to Kadungalloor | 7/27/01 | 10:30 to 10:15 a m | Red |
| 14 | Ernakulam | Mukkannur | | 7/27/01 | | Red |
| 15 | Ernakulam | Kottuvalli | | 7/27/01 | 1 p m | Red |
| 16 | Ernakulam | Parur | Kadungallur | 7/27/01 | | Red |
| 17 | Ernakulam | Muvattupuzha | puthuppadi | 7/27/01 | | Red |
| 18 | Ernakulam | Nedumbassery | Avanamkodu | 7/27/01 | | Red |
| 19 | Ernakulam | Kaitharam | | 7/27/01 | 1 p m | Red |
| 20 | Ernakulam | Aaluva | Chenganodu | 7/27/01 | 10:30 a m | Red |
| 21 | Ernakulam | Muvattupuzha | Mazhuvannur | 7/28/01 | Night | Yellow |
| 22 | Ernakulam | Alwaye | Edayar, Panayikulam, Valluvalli, Kongorppally | 7/28/01 | 8 a m | Red |
| 23 | Ernakulam | Perumbavur | Vazhakkulam | 7/28/01 | 10 to 10:15 | Red |
| 24 | Ernakulam | Nedumbassery | | 7/28/01 | | Red and Yellow |
| 25 | Ernakulam | Chengamanadu | | 7/28/01 | | Red |
| 26 | Ernakulam | Kochi | Fortkochi | 7/28/01 | 6 to 7 a m | Red |
| 27 | Ernakulam | Kochi | Mattamcherri | 7/28/01 | 6 to 7 a m | Red |
| 28 | Ernakulam | Kochi | Pallurithi | 7/28/01 | 6 to 7 a m | Red |
| 29 | Ernakulam | Mulamthurithi | Karavatte | 7/28/01 | | Red |
| 30 | Ernakulam | Kezhmadu | Chundi, Chunangamveli | 7/28/01 | | Red |
| 31 | Ernakulam | Kochi | Kadavanthra | 8/8/01 | 11 a m | Red |
| 32 | Ernakulam | Nedumbassery | chengamnad, Kulavankunnu | 8/13/01 | 7:30 a m | Red |
| 33 | Ernakulam | Angamali | Aiyampuzha, Karukutty | 8/13/01 | | Red |
| 34 | Ernakulam | Malayattoor | Thottukavala | 8/13/01 | 7:30 a m | Red |
| 35 | Ernakulam | Kochi | Ponnurunthi | 9/18/01 | 7:30 a m | Red |
| 36 | Idukki | Vannappuram | Killippara | 7/25/01 | night | Red |
| 37 | Idukki | Adimali | Machiplavu | 7/28/01 | | Deep yellow & yellow red |
| 38 | Idukki | Kanjar | | 7/28/01 | | Red |
| 39 | Idukki | Thodupuzha | | 7/28/01 | | Yellow |
| 40 | Idukki | Nariyampara | | 7/28/01 | | Red |
| 41 | Idukki | Deviyar | | 7/28/01 | | Red |
| 42 | Idukki | Vellathuval | Panar | 7/31/01 | | Red |
| 43 | Idukki | Erattayar | Natthukallu | 8/1/01 | | Red |

| No. | District | Place | Location | Date | Approximate time | Available description |
|---|---|---|---|---|---|---|
| 44 | Idukki | Kattappana | Vallakkadavu | 8/1/01 | | Red |
| 45 | Idukki | Kattappana | | 8/2/01 | | Black & red |
| 46 | Idukki | Kattappana | Irattayar | 8/4/01 | | Red |
| 47 | Kannur | Kuttuparambu | Thokkilangadi | 7/26/01 | early morning, 15 minutes | Light red |
| 48 | Kannur | Peringom | | 7/31/01 | night | Red |
| 49 | Kannur | Panoor | Chendayat | 9/21/01 | | Red |
| 50 | Kannur | Pallikunnu | Ramatheru, Kanathur, | 9/23/01 | 8:30 a m | Red |
| 51 | Kannur | Pallikunnu | Puthiyatheru | 9/23/01 | 8:30 a m | Red |
| 52 | Kannur | Edakkad | Koshormoola | 9/23/01 | 6 to 10 a m | Red |
| 53 | Kannur | Pappinisseri | Keecheri | 9/23/01 | 6 to 10 a m | Red |
| 54 | Kasaragode | Madikai | Malappacherry | 8/1/01 | | Red |
| 55 | Kollam | Sooranad | | 7/28/01 | | Red |
| 56 | Kollam | Pathanapuram | | 7/28/01 | | Red |
| 57 | Kollam | Mavila | | 8/26/01 | | |
| 58 | Kollam | Pattazhi | | 8/27/01 | | |
| 59 | Kottayam | Changanacherry | Morkulangara | 7/25/01 | 8:45 a m | Red |
| 60 | Kottayam | Changanacherry | Puzhavathu | 7/25/01 | Evening | |
| 61 | Kottayam | Puthuppally | Thrikkothamangalam | 7/27/01 | 8:30 to 9 a m | Red |
| 62 | Kottayam | Kanjirappalli | Chenappady | 7/27/01 | 1:30 to 1:40 p m | Red |
| 63 | Kottayam | Pala | Kadanadu | 7/27/01 | 5 p m | Red, white turbid |
| 64 | Kottayam | Karukachal | Edayappara | 7/28/01 | Evening | Deep yellow color |
| 65 | Kottayam | Changanacherry | Poovam | 7/28/01 | 10:30 a m | Red |
| 66 | Kottayam | Pambadi | Vazhoor | 7/28/01 | morning | Red |
| 67 | Kottayam | Ammancherri | | 7/28/01 | 6 p m | Yellow |
| 68 | Kottayam | Neendoor | Pravattom | 7/28/01 | morning | Red |
| 69 | Kottayam | Changanacherry | Vazhappally | 7/28/01 | | Red |
| 70 | Kottayam | Vechoochira | Mannadishala | 7/30/01 | | Red |
| 71 | Kottayam | Ayirur | Edappavoor | 7/30/01 | | Red |
| 72 | Kottayam | | Planghaman | 7/30/01 | | Red |
| 73 | Kottayam | | Pullappram | 7/30/01 | | Red |
| 74 | Kottayam | Manimala | Market Jn. | 8/2/01 | 10 a m | Yellow |
| 75 | Kottayam | Elikkulam Panchayat | Panamattom | 8/2/01 | | Red |
| 76 | Kottayam | Manimala | near Mini Indust. Estate | 8/4/01 | 11 a m | red |
| 77 | Kottayam | Manimala | | 8/5/01 | | Red |
| 78 | Kottayam | Erumali | Thumarampara | 8/5/01 | morning | Yellow |
| 79 | Kottayam | Mundakkayam | Punchavayal | 8/5/01 | | yellow |
| 80 | Kottayam | Changanacherry | Morkulangara | 8/14/01 | 10 a m | Red |
| 81 | Kottayam | Elikkulam Panchayat | Panamattom | 8/15/01 | | |
| 82 | Kottayam | Ettumanoor | Pattithanam | 8/19/01 | | Red, yellow |
| 83 | Kottayam | Kanjirappalli | Kappad | 9/23/01 | 8 to 8:30 a m | Red |
| 84 | Kottayam | Erumeli | Erumeli | 9/23/01 | | |
| 85 | Kozhikode | Malapparambu | | 8/4/01 | | Red |
| 86 | Malappuram | Vengara | | 7/28/01 | | |
| 87 | Malappuram | Thirur | Menathur angadi | 8/4/01 | 7:30 a m | Blue |
| 88 | Palakkad | Ongallur | Vadanamkurichy | 7/27/01 | night | Red |
| 89 | Palakkad | Kizhakkancheri | | 7/27/01 | night | Red |
| 90 | Palakkad | Vadakkancheri | | 7/27/01 | night | Red |
| 91 | Palakkad | Kannambra | | 7/27/01 | night | Red |
| 92 | Pathanamthitta | Valamchuzhi | | 7/27/01 | bet. 9:30 & 10:30 a m | Light red |
| 93 | Pathanamthitta | Valamchuzhi | | 7/27/01 | 1 p m | Red |

| No. | District | Place | Location | Date | Approximate time | Available description |
|---|---|---|---|---|---|---|
| 94 | Pathanamthitta | Chittar | Aarattukudukka | 7/27/01 | early morning | Yellow |
| 95 | Pathanamthitta | Karimpanakuzhi | | 7/28/01 | | Red |
| 96 | Pathanamthitta | Kumbuzha | | 7/28/01 | | Red |
| 97 | Pathanamthitta | Chennerkkara | Murippara | 7/28/01 | | Red |
| 98 | Pathanamthitta | Adoor | Kilivayal | 7/28/01 | | Red |
| 99 | Pathanamthitta | Adoor | Munnalam | 7/28/01 | | Red |
| 100 | Pathanamthitta | Adoor | Karuvatta | 7/28/01 | | Red |
| 101 | Pathanamthitta | Kidangur | Kotta | 7/28/01 | | Red |
| 102 | Pathanamthitta | Panthalam | Kulanda | 7/28/01 | | Red |
| 103 | Pathanamthitta | | Kaippuzha | 7/28/01 | | Red |
| 104 | Pathanamthitta | Panthalam | Kadakkadu | 7/28/01 | | Red |
| 105 | Pathanamthitta | | Mangaram | 7/28/01 | | Red |
| 106 | Pathanamthitta | | Vallikkod | 7/28/01 | | Red |
| 107 | Pathanamthitta | | Thrikkovil | 7/28/01 | | Red |
| 108 | Pathanamthitta | Thiruvalla | Kuttur | 7/28/01 | | Red and yellow hailstones |
| 109 | Pathanamthitta | Thiruvalla | Peringara | 7/28/01 | Morning | Red |
| 110 | Pathanamthitta | Ranni | Palachuvadu | 7/29/01 | 9:30 a m | Red and Yellow |
| 111 | Pathanamthitta | Murinjakal | Mlamthadam | 8/1/01 | morning for 5 minutes | Red |
| 112 | Pathanamthitta | Kozhencherrry | | 8/1/01 | | Black rain |
| 113 | Pathanamthitta | Vallikode | Vellappara Colony | 8/1/01 | | |
| 114 | Pathanamthitta | Kadammanitta | | 8/1/01 | | |
| 115 | Pathanamthitta | Ranni | Vechoochira | 8/4/01 | | |
| 116 | Thiruvananthapuram | Vithura | Kallar | 7/31/01 | | Red |
| 117 | Trichur | Ponkunnam | Elangulam | 7/30/01 | | Both Light black and red colored types |
| 118 | Trichur | Gruvayoor | Mammiyoor | 9/23/01 | 9:30 a m | Red |
| 119 | Trichur | Anthikad | | 9/23/01 | 9:30 a m | |
| 120 | Trichur | Irinjalakuda | Arippalam | 9/23/01 | 9:30 a m | Red |
| 121 | Trichur | Puttanpedika | Tattaadi | 9/23/01 | 9:30 a m | Pink |
| 122 | Wayanad | Padinharathara | Dam site | 8/1/01 | 10:15 to 10:40 a m | Red |
| 123 | Wayanad | Ambalavayl | Edakkal | 8/4/01 | 12 noon | Red |
| 124 | Wayanad | Thalappuzha | Kannothmala | 8/5/01 | | Red |